\documentclass[aps,pra,twocolumn,superscriptaddress,showkeys,amsmath,amssymb]{revtex4-1}		
\usepackage{graphicx}% Include figure files
\usepackage{dcolumn}% Align table columns on decimal point
\usepackage{bm}% bold math
\usepackage{hyperref}% add hypertext capabilities
\usepackage{subfig}
\usepackage[font=small,labelfont=bf,format=plain,justification=centerlast,labelsep=period]{caption}
\usepackage[usenames, dvipsnames]{color}

\usepackage{verbatim}
\usepackage[normalem]{ulem}

\begin{document}

\preprint{APS/123-QED}

\title{Damped Casimir radiation 
and photon correlation measurements}

\author{R. Rom\'an-Ancheyta}
\email{ancheyta6@gmail.com}
\affiliation{Instituto de Ciencias F\'isicas,
Universidad Nacional Aut\'onoma de M\'exico, Apartado Postal 48-3, 62251 Cuernavaca, Morelos, M\'exico}
\author{O. de los Santos-S\'anchez}
\email{octavio.desantos@gmail.com}
\affiliation{Instituto de Ciencias F\'isicas,
Universidad Nacional Aut\'onoma de M\'exico, Apartado Postal 48-3, 62251 Cuernavaca, Morelos, M\'exico}
\author{C. Gonz\'alez-Guti\'errez}
\email{carlosgg04@gmail.com}
\affiliation{Instituto de Ciencias F\'isicas,
Universidad Nacional Aut\'onoma de M\'exico, Apartado Postal 48-3, 62251 Cuernavaca, Morelos, M\'exico}
\date{\today}

\begin{abstract}
An effective toy model for an ideal one-dimensional nonstationary cavity is taken to be the starting point to derive a fitting markovian master equation for the corresponding leaky cavity. In the regime where the generation of photons via the dynamical Casimir effect is bounded, the master equation thus constructed allows us to investigate the effects of decoherence on the average number of Casimir photons and their quantum fluctuations through the second-order correlation function. 

%\begin{description}
%\item[PACS number(s)]
%03.70.+k, 42.50.Pq, 42.82.Et, 42.50.Lc
%\end{description}
\end{abstract}

\maketitle

\section{Introduction}

Cavity dynamical Casimir effect (DCE) is a
fascina\-ting quan\-tum mecha\-nical phenomenon
in which real photons can be created out of
vacuum fluctuations, via parame\-tric amplification,
as a consequence of nonadiabatic changes
in the time-dependent electromagnetic 
cavity boundary conditions~\cite{DodonovReview}.
Theoretically predicted by Moore in the 
70's~\cite{moore1970},
the DCE was demonstrated experimentally,
more than forty years later, by using a 
superconducting Josephson metamaterial as 
a surrogate for a fast 
(a significant fraction of the speed of light) 
oscillating cavity mirror~\cite{Lahteenmaki12032013}.
Besides being an important result 
from the fundamental point of view of  
quantum field theory,
the aforesaid phenomenon has also been investigated in a number of contexts such as 
trapped ions~\cite{Nils},
quantum refri\-ge\-ra\-tors~\cite{JPaz},
Kerr media~\cite{ricardo2017},
and, still more recently, in 
stochastic systems~\cite{ricardoguias};
furthermore, it has found interesting applications
in circuit quantum electrodynamics~\cite{RevModPhysFNori}
where entangled artificial atoms~\cite{Solano1} and 
Gaussian boson samplers~\cite{Borja1} can be realized. 

On the other hand, the issue of decoherence has also attracted a great deal of interest in recent times since, as known, a given physical system cannot be completely isolated from its surroundings; indeed, a more realistic scenario has to take into account the effects of loss of quantum coherence that stem from being oblivious of the environmental influences. In this regard, of particular interest to us is the fact that if one wants to detect and correlate Casimir photons, it is essential to consider the unavoidable interaction with their environment. Along this line of research, there have been several attempts to incorporate decoherence and  energy losses in nonstationary cavities including either amplitude~\cite{Dodonovlosses} or phase~\cite{Ralfdecohe} damping; and a rather cumbersome time dependent master equation has already been derived from first principles as asserted in Ref.~\cite{Soffleaky}. However, despite all these efforts, a consensus about how to properly analyze the effects of dissipation on the DCE has not yet reached. 
\begin{center}
\begin{figure}[b!]
\includegraphics[width=6.5cm, height=2.5cm]{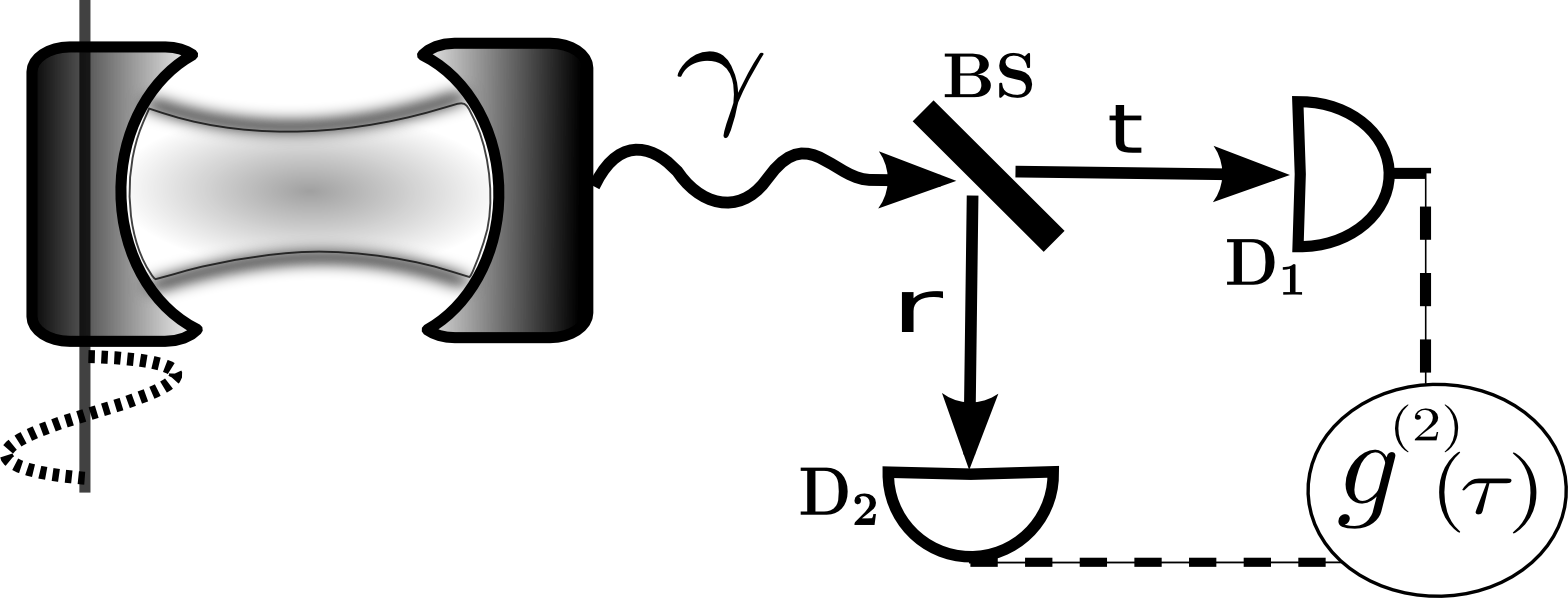} 
\caption{Optical scheme for detecting Casimir radiation.
(a) Nonstationary leaky cavity in which the DCE  
is manifested. (b) Cre\-a\-ted photons that leak out the
cavity are detected and correlated by a Hanbury-Brown and
Twiss apparatus composed of a beam splitter, BS, and two photodetectors, $D_1$ and $D_2$, to perform the measurement of $g^{(2)}(\tau)$.}
\label{experimental_setup}
\end{figure}
\end{center}
This work is in keeping with the spirit of putting forward an algebraic toy model capable of describing, in an effective manner, decoherence effects in the process of creating and correlating photons in the DCE, a proposed model that is considered to be justified only under certain environmental and system conditions. More precisely, we restrict ourselves to a parameter regime in which the generation of photons remains bounded so as to be able to derive a fitting markovian master equation for the reduced density operator of the dynamical cavity viewed as an open quantum system. The optical scheme of the process it seeks to describe is sketched in Fig.~\ref{experimental_setup}. According to the figure, a nonstationary leaky cavity takes on the role of our central system from which Casimir photons are created and their quantum fluctuations are analyzed in terms of the second-order correlation (coheren\-ce) function by means of a standard Hanbury-Brown and Twiss intensity interferometer. The details of the effective model and the precise parameter settings within which we shall be focused on will be given below.

The simplest effective
Hamiltonian des\-cribing, in the Schr\"odinger 
picture, the dynamics of an electromagnetic
field inside a lossless one-dimensional 
nonstationary single-mode cavity, whose instantaneous frequency  
$\omega(t)=\omega_0[1+\epsilon\sin(\nu t)]$ follows from the time-dependent geometry of the system, 
is given by \cite{Law1,Dalvit2011,DodonovOneAtom,DodonovTwoAtoms} ($\hbar=1$):
\begin{equation}\label{eq:hamiltonian}
\hat{H}_\mathrm{eff}=(\epsilon\omega_0/4)(\hat a^{\dagger 2}+\hat a^2)+(K/2)\hat n,
\end{equation}
where $\omega_0$ is the fundamental frequency of the cavity, and $\epsilon$ ($\nu$) is 
the amplitude (frequency) modulation. 
This time-independent Hamiltonian is an effective algebraic model obtained under quasi-resonant conditions, i.e., 
$\nu=2\omega_0+K$, with $K$ being a small 
frequency shift, and is written in a rotated referen\-ce frame in which 
the rotating-wave approximation (RWA) is also 
employed~\cite{DodonovOneAtom,DodonovTwoAtoms}. The SU(1,1) symmetry of $\hat{H}_\mathrm{eff}$, generated by the set of operators $\{ \hat{n}, \hat{a}^{\dagger 2}, \hat{a}^{2} \}$, enables us to obtain, by making use of standard Lie algebraic methods, the following expression for the expectation value of the number of photons generated from the vacuum state~\cite{Ban93}:
\begin{equation}\label{empty_cavity}
\langle \hat n\rangle=\sinh^2(\eta\epsilon\omega_0t/2)/\eta^2,
\end{equation}
with $\eta=\sqrt{1-(K/\epsilon\omega_0)^2}$. Hence, depending on whether the ratio $K/\epsilon\omega_0$ is smaller or greater than unity, one can identify a twofold behavior of photon generation ranging from the exponential growth ($K/\epsilon\omega_0<1$) to the bounded oscillatory regime ($K/\epsilon\omega_0>1$), in which case the argument of the hyperbolic function becomes imaginary so that the replacement $\eta\rightarrow\tilde{\eta}=\sqrt{(K/\epsilon\omega_0)^2-1}$ follows; this crossover has recently been referred to as a 
metal-insulator phase transition~\cite{ricardoguias}.

Returning to the subject of considering the nonstationary cavity as an open system, an ansatz for the corresponding phenomenological master equation at zero temperature is considered to have the following structure \cite{Dodonovlosses}: $\mathrm{d}\hat{\rho}/\mathrm{d}t =-i[\hat{H}_\mathrm{eff},\hat{\rho}]+\kappa \mathcal{L}[\hat{a}]\hat{\rho}$, where $\kappa$ is the decay rate, which is inversely proportional to the quality factor of the cavity, and the generator $\mathcal{L}[x]$ is such that $\mathcal{L}[x]\hat{\rho}\equiv 2x\hat{\rho} x^\dagger-x^\dagger x\hat{\rho}-\hat{\rho} x^\dagger x$, with $\hat{\rho}$ being the reduced system density operator. On the basis of this master  equation, we arrive at the modified version of \eqref{empty_cavity}:
\begin{align}\label{n_average_pheno}
\langle \hat{n}\rangle=&
-2\langle \hat n\rangle_{st}^{ph}e^{-2\kappa t}\big[\sinh^2(\eta\epsilon\omega_0 t/2)+{1}/{2}
\nonumber\\
&+({\kappa}/{\eta\epsilon\omega_0})\sinh(\eta\epsilon\omega_0 t)\big]+
\langle \hat n\rangle_{st}^{ph},
\end{align}
where $\langle \hat n\rangle_{st}^{ph}=\frac{1}{2}
[(2\kappa/\epsilon\omega_0)^2-\eta^2]^{-1}$ 
is the steady state average of the photon number provided that
$2\kappa>\eta\epsilon\omega_0$. In the resonant case, $K=0$, the result of Ref. \cite{Dodonovlosses} is recovered from \eqref{n_average_pheno} and, again, a photon-number exponential growth is obtained as long as the amplitude modulation surpasses the rate at which the system decays. On the other hand, it was shown in \cite{Ralfdecohe} that depha\-sing effects themselves, $\propto \mathcal{L}[\hat{n}]\hat{\rho}$, tend to only slow down the photon generation rate. In this approach, however, the lack of knowledge about the explicit form of the system's steady state makes it difficult, for instance, to examine analytically the statistical behavior of the outgoing photons via the second order correlation function involving the use of the known quantum regression formula \cite{carmichael}. Further drawbacks of the phenomenological treatment have already been discussed in \cite{Soffleaky}.

Section \ref{sec:2} outlines the derivation of the microscopic master equation on the basis of the Born and Markov approximations, an approach that is suitable for the description of dissipation in the DCE evolving within the bounded regime of photon generation. Having determined the steady state limit of our system, we proceed, in section \ref{sec:3}, to the description of the outcome of the proposed master equation reflected upon the average photon number and the intensity correlation of two created photons. And finally, some conclusions are given in section \ref{sec:4}. 

%%%%%%%%%%%%%%%%%%%%%%%%%%%%%%%%%%%%%%%%%%%%%%%%%%%%%%%%%%%
\section{Microscopic master equation} \label{sec:2}

Let the time-independent effective Hamiltonian (\ref{eq:hamiltonian}) be our starting point. This Hamiltonian can easily be diagonalized by making use of the squeeze operator $\hat{S}(r) = \exp[ r(\hat{a}^{2}-\hat{a}^{\dagger 2})/4 ] = e^{ z\hat{a}^{\dagger 2}/2} e^{\frac{\beta}{2}\left(\hat{n}+\frac{1}{2}\right)}e^{-z \hat{a}^{2}/2 }$ through the unitary transformation $\hat{S}(r)\hat{H}_\mathrm{eff}\hat{S}^{\dagger}(r)$;  here, the particular choice of the parameter 
$r=\frac{1}{2}\ln[{(K+\epsilon \omega_0)}/{(K-\epsilon \omega_0)}]$
guarantees the proper diagonalization process provided that the inequality $K/\epsilon \omega_0>1$ holds, and the remaining parameters $z=-\tanh (r/2)$ and $\beta = \ln (1-z^{2})$ are stated by disentangling the exponential. So, via this reframed system, it is found that the corresponding eigenenergies and eigenstates of our effective Hamiltonian are, respectively, given by
\begin{align} 
2E_{n}  &=\epsilon \omega_{0} \sqrt{(K/\epsilon \omega_{0})^2-1}(n+{1}/{2})-{K}/{2}, 
\label{eq:eigenenergies}\\
|r ,n\rangle  &=  \hat{S}(r)|n\rangle, \label{eq:eigenstates}
\end{align}
where $n=0,1,2,\ldots$, and the latter turn out to be the so-called squeezed number states \cite{nieto}. 

Based upon the aforesaid unitary operator, we find it convenient to introduce the so-termed pseudo-annihilation (creation) operator, $\hat{b}$ ($\hat{b}^{\dagger}$), firstly introduced by Yuen \cite{yuen} in his work on two-photon coherent states, defined by
\begin{subequations} 
\begin{align} 
\hat{b} & =  \hat{S}(r) \hat{a} \hat{S}^{\dagger} (r)  =  \cosh \left({r}/{2} \right)\hat{a}+\sinh \left( {r}/{2} \right)\hat{a}^{\dagger}, \label{eq:Aop} \\
\hat{b}^{\dagger} & =  \hat{S}(r) \hat{a}^{\dagger} \hat{S}^{\dagger} (r)  =  \cosh \left({r}/{2} \right)\hat{a}^{\dagger}+\sinh \left({r}/{2} \right)\hat{a}, \label{eq:Adop}
\end{align}
\end{subequations}
which is nothing but a Bogoliubov transformation that generates $\hat{b}$ and $\hat{b}^{\dagger}$ from the standard operators $\hat{a}$ and $\hat{a}^{\dagger}$, thereby preserving the commutator $[\hat{b},\hat{b}^{\dagger}]=1$.  Thus, in this representation, one is able to obtain the following diagonal form of the system Hamiltonian (\ref{eq:hamiltonian})
\begin{equation}
\hat{H}_{S}= \Omega \big(\hat{b}^{\dagger}\hat{b}+{1}/{2}\big)-{K}/{4}, 
\label{eq:pseudoh}
\end{equation}
with the identification 
$\Omega = \tilde{\eta}\epsilon\omega_0/2$; note that the states given by (\ref{eq:eigenstates}) are indeed eigenstates of the operator $\hat{b}^{\dagger} \hat{b}$, also called quasi-number operator~\cite{mandel}. To be more precise, in this algebraic scheme, the action of $\hat{b}$ and $\hat{b}^{\dagger}$ upon the eigenstates of the system, the squeezed number states, is reminiscent of that of $\hat{a}$ and $\hat{a}^{\dagger}$ upon the Fock states in such a way that $\hat{b}|r,n\rangle=\sqrt{n} |r,n-1\rangle$ and $\hat{b}^{\dagger} |r,n\rangle = \sqrt{n+1} |r,n+1\rangle$ \cite{mandel}. I.e., such operators lower and raise one excitation by changing the number of quanta in $\pm 1$, thereby  connecting transitions between adjacent energy levels within the squeezed-number-state basis. Indeed, these operators can also be regarded as the actual eigenoperators of the system Hamiltonian in 
the sense that they obey the commutation relationships:
\begin{equation}
[\hat H_{S}, \hat{b} ]  =  -\Omega \hat{b}, \qquad [\hat H_{S}, \hat{b}^{\dagger }]  =  \Omega \hat{b}^{\dagger}.
\label{eq:iegenops}
\end{equation}
This fact motivates us to consider the possibility of deriving a fitting, albeit approximate, master equation, in the weak-coupling and Markovian regimes, in order to explore the damped dynamics of the system described by the effective  Hamiltonian (\ref{eq:hamiltonian}) but from a different algebraic point of view. That is, in such a representation, given by the pseudo-harmonic oscillator outlined above, we should be able to establish the proper master equation in a way such that the dissipative part of the evolution be modeled in terms of the actual system's eigenoperators. To this end, let the Hamiltonian describing our quadratic oscillator as an open system be structured as follows: 
$\hat{H}=\hat{H}_{S}+\hat{H}_{E}+ \hat{H}_{SE}$,
where $\hat{H}_{S}$ corresponds to the unperturbed central system (S), $\hat{H}_{E}$ is the free Hamiltonian of the environment (E), and $\hat{H}_{SE}$ represents the interplay between them. Proceeding in the customary fashion, let the environment of the central system be modeled as a bath of harmonic oscillators, i.e., $\hat{H}_{E}= \sum_{k} \omega_{k} \hat{B}_{k}^{\dagger}\hat{B}_{k}$, with $\omega_{k}$ being the frequency of the {\it k}-th oscillator, and their interaction be governed by the following Hamiltonian in the Schr\"odinger picture
\begin{equation}
\hat{H}_{SE} = (\hat{a}+\hat{a}^{\dagger})\small{\sum}_{k}g_{k}(\hat{B}_{k}+\hat{B}_{k}^{\dagger}),\label{eq:interaction}
\end{equation}
which is taken to be linear in both the cavity field and the environment amplitudes, $\hat{B}_{k}$ ($\hat{B}^{\dagger}_{k}$) is the annihilation (creation) operator within the {\it k}-th mode, and the $g_{k}$'s are the coupling parameters;  \textcolor{black}{incidentally, this kind of system-environment interaction resembles the multimode coupling Hamiltonian model proposed in Ref.~\cite{Soffleaky}, where the authors attempt to describe a leaky-cavity configuration in which a dispersive mirror is inserted into a larger ideal nonstationary cavity that, in turn, plays the role of a reservoir.} So, Hamiltonian (\ref{eq:interaction}), written in terms of the na\-tu\-ral variables of the field, $\hat{a}$ and $\hat{a}^{\dagger}$, can be recast in the pseudo-harmonic-oscillator representation by inversion of (\ref{eq:Aop}) and (\ref{eq:Adop}). So, in the interaction picture generated by $H_{S}+H_{E}$, with $H_{S}$ being taken to be (\ref{eq:pseudoh}), we get
\begin{equation}
\tilde{H}_{SE}(t) =(\hat{b}e^{-i\Omega t}+\hat{b}^{\dagger}e^{i\Omega t})\small{\sum}_{k}g_{k}(r)
(\hat{B}_{k}e^{-i\omega_{k}t}+\hat{B}_{k}^{\dagger}e^{i\omega_{k}t}), \label{eq:Hset}
\end{equation}
where the coupling constants are now construed as being dependent on the squeezing parameter, that is to say,  $g_{k}(r)=e^{-r/2}g_{k}$. Based upon the Born and Markov approximations, and within the framework of the master equation approach, one is able to establish the following equation for the reduced density operator associated with the system at hand which makes no restriction on the precise interaction between the latter and its surroundings  \cite{gardiner}
\begin{equation}
\dot{\tilde{\rho}}(t) = -\int_{0}^{\infty} d\tau Tr_{E} \big \{ \big[ \tilde{H}_{SE}(t), \big [\tilde{H}_{SE}(t-\tau),\tilde{\rho}(t) \otimes \rho_{E}\big]\big] \big \},
\label{eq:master}
\end{equation}
where the tilde over the density operator means that it is in the interaction picture, $Tr_{E}$ indicates the trace over the environment variables, and $\rho_{E}$ represents the state of the environment which, according to the Born approximation, is taken to be constant in its evolution ($\rho_{E}(t)\approx \rho_{E}(0)=\rho_{E}$) and determined by the Boltzman distribution $\rho_{E} = e^{-\hat H_{E}/k_{B}T}/Tr \{e^{-\hat{H}_{E}/k_{B}T}\}$.
Substitution of (\ref{eq:Hset}) into (\ref{eq:master}), application of the RWA to the equation thus obtained, and going back to the Shr\"odinger representation, leads us, after some algebra, to a Lindblad master equation describing the damped dynamics of the system interacting with a bath of harmonic oscillators in thermal equilibrium at $T$ temperature:
\begin{equation} 
\frac{\mathrm{d}\hat{\rho}}{\mathrm{d}t} 
=  -i [\hat{H}_{S},\hat{\rho}]+ \gamma_{r} \left ( N_{\Omega}+1 \right) \mathcal{L} [\hat{b}]\hat{\rho} +\gamma_{r} N_{\Omega} \mathcal{L} [\hat{b}^{\dagger}] \hat{\rho} .
\label{eq:mastert}
\end{equation}
Here, $\gamma_{r}=e^{-r}{\gamma}$ is the overall damping rate, where we have let ${\gamma}=\pi h(\Omega) |g(\Omega)|^{2}$ be, approximately, a constant quantity, provided of course that a spectrally flat environment is taken into consideration, with $h(\Omega)$ and $g(\Omega)$ being, respectively, the density of states and the system-environment coupling at $\Omega$; and $ N_{\Omega}=1/(e^{\hbar \Omega/k_{B}T}-1)$ is the average number of thermal photons in the reservoir at the aforesaid frequency. 

The dependency of the resulting master equation upon the squeezing parameter $r$ is apparent from the viewpoint of the algebraic scheme we are working on: each set of parameters $\{K, \epsilon, \omega_{0} \}$, in terms of which the squeezing one is set down, is thought of as defining, correspondingly, a different oscillator system described by the algebraic Hamiltonian (\ref{eq:pseudoh}), whose energy spectrum, albeit equally spaced for a given value of the parameters involved (see Eq. (\ref{eq:eigenenergies})), displays an overall dependency on the ratio $K/\epsilon\omega_{0}$ with a tendency to merge at $K/\epsilon\omega_{0}=1$. So, for each value of this ratio, we have specific pseudo operators $\{\hat{b}, \hat{b}^{\dagger}\}$ describing the allowed transitions induced by the environment that take place at the time scales $\gamma_{r}^{-1}$ and at the specific transition frequency $\Omega$ between the energy levels involved in accord with the commutation relations (\ref{eq:iegenops}). 
It is also worth commenting that taking advantage of the algebraic scheme in terms of which it is written down, Eq. (\ref{eq:mastert}) can be solved by using, for example, the standard technique based upon superoperators (see, for instance, Ref.~\cite{moya}) allowing us to confirm that in the asymptotic limit $t\to \infty$, the reduced density operator approaches that of the squeezed thermal state, namely, $\lim_{t\rightarrow\infty}\hat{\rho}=(1+ N_{\Omega})^{-1} \sum_{n} \left(\frac{ N_{\Omega}}{1+ N_{\Omega}}\right)^{n} |r,n \rangle \langle r, n|$, and, therefore, the steady state at zero temperature ($ N_{\Omega}=0$) becomes precisely the squeezed vacuum state, $\hat{\rho}(t\to \infty) \to |r,0\rangle \langle r,0|$, which clearly corresponds to the state of minimum energy (the ground state) of the system (see Eq. (\ref{eq:eigenstates})).
\textcolor{black}{We note in passing that \eqref{eq:mastert}, wri\-tten in terms of $\hat{a}$ and $\hat{a}^\dagger$, is somewhat similar to the one obtained by considering a bosonic system coupled to a phase-sensitive reservoir~\cite{ekert} and, additionally, bears some resemblance to the master equation derived in Ref.~\cite{Soffleaky}.}
%%%%%%%%%%%%%%%%%%%%%%%%%%%%%%%%%%%%%%%%%%%%%%%%%%%%%%%%%%%
\section{Generation and correlation of photons} \label{sec:3}

\begin{center}
\begin{figure}[t!]
\includegraphics[width=7cm, height=5cm]{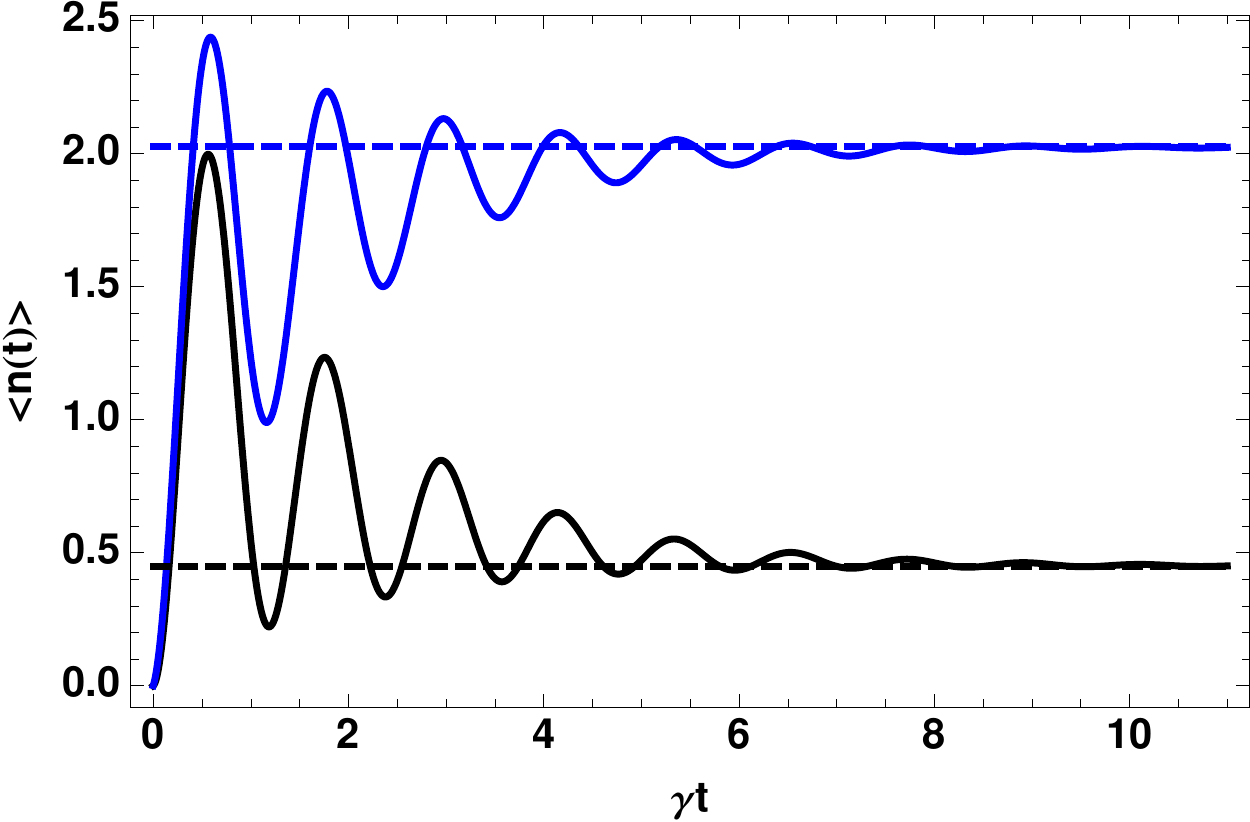} 
\caption{Plot of the average photon number (\ref{eq:nprom}) as a function of the scaled time $\gamma t$. One can visualize both the transient period of the system and the steady state limit of it for $ N_{\Omega}=0$ (black line) and $ N_{\Omega} \neq 0$ (blue line) such that $\hbar K/k_{B}T=3$, $K/\gamma =10$, and $\epsilon \omega_{0}/K=0.85$. }
\label{fig:generation}
\end{figure}
\end{center}

Let us now discuss some results concerning the damped evolution of the average photon number and the second-order intensity correlation function. Firstly, to evaluate the expectation value of the number operator in the current representation, namely, $\langle \hat{n}\rangle  =  \cosh(r) \langle \hat{b}^{\dagger}\hat{b}\rangle-\frac{1}{2}\sinh(r)[\langle \hat{b}^{\dagger 2}\rangle+ \langle\hat{b}^{2}\rangle]+\sinh^{2}( r/2 )$, we find it convenient to establish the following equations of motion for the its constituents:
\begin{subequations}
\begin{align}
\mathrm{d}\langle \hat{b}^{\dagger}\hat{b} \rangle/ \mathrm{d}t & =  -2 \gamma_{r} \langle \hat{b}^{\dagger} \hat{b} \rangle+2 \gamma_{r} N_{\Omega}, \\
\mathrm{d}\langle \hat{b}^{2} \rangle /\mathrm{d}t & =  -(2i\Omega+2\gamma_{r}) \langle \hat{b}^{2} \rangle,
\end{align}
\end{subequations}
which follows from applying $\langle \dot{ \hat{O}} \rangle=Tr \{ \dot{\hat{\rho}} \hat{O} \} $ with the help of (\ref{eq:mastert}). The solution to these equations is quite straightforward and a closed-form expression for the average photon number can be found when the state of the system at the initial moment of time is the vacuum, $\hat{\rho}(0)=|0\rangle \langle 0|$, represented in the squeezed-number-state eigenbasis (\ref{eq:eigenstates}). Thus, the sought result turns out to be
\begin{align}
\langle \hat{n} (t) \rangle & =  
e^{-2\gamma_{r}t} \sin^{2}\left(\tilde{\eta}\epsilon\omega_0 t/2 \right)/\tilde{\eta}^2 
 +(1-e^{-2\gamma_{r}t})\langle\hat{n}\rangle_{st}, \label{eq:nprom}
\end{align}
in which one is able to distinguish clear-cut limits being represented in Fig.~\ref{fig:generation}: (i) on the one hand, we identify the undamped case ($\gamma_{r}=0$) that corresponds to the well-known result regarding the oscillatory regime of photon generation; (ii) on the other hand, once the transient evolution has elapsed, it is found that the steady state limit, $\lim_{t \to \infty} \langle \hat{n}(t) \rangle =\langle \hat{n} \rangle_{st}= \langle \hat{n} \rangle_{st,0}(1+2N_{\Omega})+ N_{\Omega}$, corresponds to the mean photon number for the squeezed thermal state (blue line), with $\langle \hat{n} \rangle_{st, 0} = \frac{1}{2}[(1-(\epsilon \omega_{0}/K)^{2})^{-1/2}-1 ]$ being the corresponding average without thermal photons, $ N_{\Omega} = 0$ (black line).
\textcolor{black}{Note that, in contradistinction to the phenomenological master equation's prediction, $\langle\hat{n}\rangle_{st}$ is independent of the rate at which the central system decays.}
\textcolor{black}{\eqref{eq:nprom}, which is one of the main results of this paper, could in principle be tested  experimentally by measuring the probability distribution of the outgoing cavity photons at different time intervals with the use of just one of the photo-detectors sketched in Fig.~\ref{experimental_setup}.}

\begin{center}
\begin{figure}[t!]
\includegraphics[width=7cm, height=5cm]{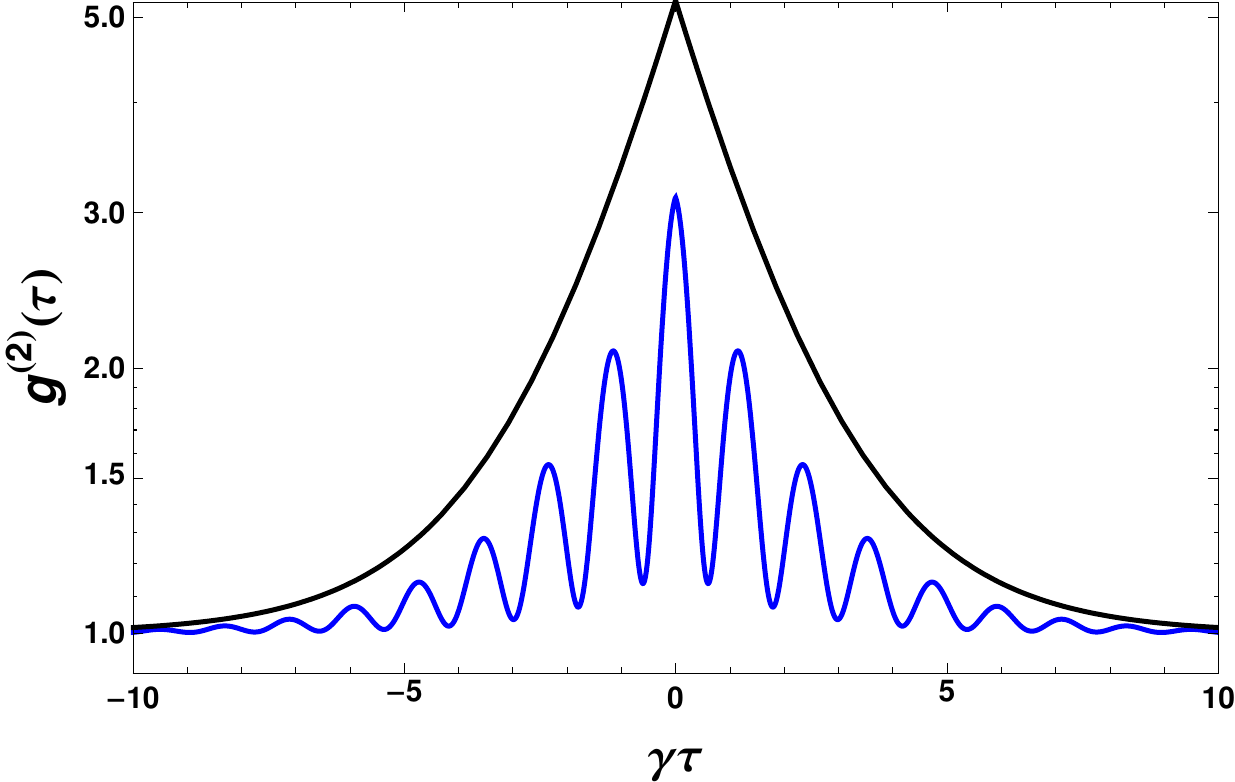} 
\caption{Normalized second-order correlation function (\ref{eq:g2}) as a function of the scaled time delay $\gamma \tau$. The set of parameters is the same as in Fig. \ref{fig:generation}.}
\label{fig:g2plot}
\end{figure}
\end{center}

As far as the intensity-intensity correlation function is concerned, we calculate the standard normalized  expression of it given by $g^{(2)}(\tau) = \langle \hat{a}^{\dagger}(0)\hat{a}^{\dagger}(\tau) \hat{a}(\tau) \hat{a}(0) \rangle/\langle \hat{a}^{\dagger} \hat{a} \rangle^{2}_{st}$.
This correlation function, which is intrinsically time-symmetric, represents a relative measure of the joint probability of detecting two photons separated by a time delay $\tau$ and allows us to discern whether two detection processes are correlated or independent of each other.
\textcolor{black}{According to the optical scheme displayed in Fig.~\ref{experimental_setup}, in order to obtain the value of $g^{(2)}(\tau)$, the outgoing generated photons are passed through a 50/50 beamsplitter, the transmitted signal registered by detector $D_1$ is multiplied with the reflected one registered by $D_2$ and, in turn, they are averaged over all the detected values~\cite{ficekbook}.} 
Applying the quantum regression formula \cite{carmichael}, together with (\ref{eq:nprom}), enables us to arrive at the expression 
\begin{equation}
g^{(2)}(\tau) = 1+e^{-2\gamma_{r}\tau} \left [ C_{1}+C_{2}\cos(2\Omega \tau) \right ],
\label{eq:g2}
\end{equation}
where we have set the constant terms
\begin{subequations}
\begin{align}
C_{1} & =  \{ (1+2\langle \hat{n} \rangle_{st,0})\left[ \langle \hat{n} \rangle_{st}- N_{\Omega} \left (\langle \hat{n} \rangle_{st}+\langle \hat{n} \rangle_{st,0}+2 \right) \right] \nonumber \\
&  +(1+2\langle \hat{n} \rangle_{st,0})^{2}(2 N_{\Omega}^{2}+ N_{\Omega}) \}/\langle \hat{n} \rangle_{st}^{2}, \\
C_{2} & =  %\left(\frac{\epsilon \omega_{0}}{2\Omega}\right)^{2}
{ N_{\Omega}\left( N_{\Omega}+1\right)}/{\langle \hat{n} \rangle^{2}_{st}}\tilde{\eta}^2. \label{eq:c2}
\end{align}
\end{subequations}
\eqref{eq:g2} behaves differently, as a function of the photo-counting delay $\tau$, as shown in Fig. \ref{fig:g2plot}, depending on whether or not thermal photons come into play. At zero temperature (black line), we see that the correlation function reveals a noticeable bunching effect for short time delays; particularly, $g^{(2)}(0)|_{N_\Omega=0}=3+\langle \hat{n}\rangle_{st,0}^{-1}$ reveals a super-thermal photon statistics behavior that is in accord with the fact that, in the DCE, photons are created in pairs~\cite{Law1}. In addition to this, when $T\neq 0$ (blue line) the oscillatory fingerprint of (\ref{eq:g2}) comes about in a series of beats at the frequency $2\Omega$ that, according to (\ref{eq:c2}), are bolstered by the thermal photons from the environment via the quadratic nature of the central system in a way such that both features reinforce each other. 
\section{Conclusions} \label{sec:4}

An effective algebraic model, viewed as an open quantum system, has been proposed as a step towards a better understanding of decoherence effects on detecting and correlating photons in a nonstationary leaky cavity in which the DCE takes place. For the model to be applicable, we have restricted ourselves to the regime within which the photon generation is bounded for a given set of system parameters, i.e., when the inequality $K/\epsilon \omega_{0}>1$ is satisfied. Based upon the markovian master equation derived under this condition, it is found that the steady state limit  of the system corresponds to the squeezed thermal state at finite temperature and, thus, to the squeezed vacuum state at zero temperature; this result, as opposed to the phenomenological treatment, allows for an explicit analysis of the outgoing photons in terms of the second-order correlation function. This last feature, besides reveling a conspicuous bunching effect, turns out to display a beating behavior that is fostered by merging the environmental thermal photons and the intrinsic quadratic character of the system. The procedure outlined in this letter can also be applied, and suitable, to pursue the investigation of another \textcolor{black}{measurements of phase-dependent quantum fluctuations} in the context of Casimir ratiadion, such as the time-dependent physical spectrum of light~\cite{Eberly77}, the spectrum of squeezing~\cite{Collett}, and the amplitude-intensity correlation function~\cite{castro2016}; the last one, unlike the $g^{(2)}(\tau)$ function, is a wave-particle correlation that can exhibit large time asymmetries~\cite{castro2017}.

\section*{Acknowledgments}
R.R.-A. and C.G.-G. thank CONACYT, Mexico, for 
financial support under Scholarships Nos. 
379732 and 385108, respectively, and DGAPA-UNAM,
Mexico, for support under Project No. IN113016.
O. de los S.-S wants to thank Professor J. R\'ecamier for his hospitality at Instituto de Ciencias
F\'isicas, UNAM.

%\bibliographystyle{ieeetr}
%\bibliographystyle{ieeetr}
%\newpage
%\bibliography{casimir_dinamico}
% Produces the bibliography via BibTeX.
%\bibliographystyle{ieeetr}
%\section*{References}
%
%\bibliography{casimir_dinamico}
% Produces the bibliography via BibTeX.
%\bibliographystyle{ieeetr}
%\bibliographystyle{unsrt}

\end{document}